\documentclass[aps,prd,preprint,floatfix,nofootinbib,showpacs,superscriptaddress]{revtex4-1}
\usepackage{amsmath,amsfonts,amssymb,bm}
\usepackage{graphicx}
\usepackage{subfigure}
\usepackage{multirow}
\usepackage{longtable,booktabs}

\usepackage{color}
\definecolor{purple}{rgb}{0.5,0,0.5}
\definecolor{blue}{rgb}{0.0,0,0.9}
\usepackage[colorlinks=true, pdfstartview=FitV, linkcolor=purple, citecolor= purple, urlcolor=blue]{hyperref}

\begin{document}


\title{Quantum Numbers of the Pentaquark States $\textrm{P}_{\textrm{c}}^{+}$ via Symmetry Analysis}

\author{Chong-yao Chen }
\affiliation{Department of Physics and State
Key Laboratory of Nuclear Physics and Technology, Peking
University, Beijing 100871, China}

\author{Muyang Chen }
\email[Present address: ]{Department of Physics, Hunan Normal University, Changsha 410081, China}
\affiliation{Department of Physics and State
Key Laboratory of Nuclear Physics and Technology, Peking
University, Beijing 100871, China}

\author{Yu-xin Liu}
\email[Corresponding author: ]{yxliu@pku.edu.cn}
\affiliation{Department of Physics and State Key Laboratory of
Nuclear Physics and Technology, Peking University, Beijing 100871,
China}
\affiliation{Collaborative Innovation Center of Quantum Matter, Beijing 100871, China}
\affiliation{Center for High Energy Physics, Peking
University, Beijing 100871, China}

\date{\today}

\begin{abstract}
We investigate the quantum numbers of the pentaquark states $\textrm{P}_{\textrm{c}}^{+}$, which are composed of four (three flavors) quarks and an antiquark, by analyzing their inherent nodal structure in this paper.
Assuming that the four quarks form a tetrahedron or a square, and the antiquark locates at the center of the four quark cluster,
we determine the nodeless structure of the states with orbital angular moment $L \leq 3$, and in turn, the accessible low-lying states.
Since the inherent nodal structure depends only on the inherent geometric symmetry,
we propose the quantum numbers $J^{P}$ of the low-lying pentaquark states $\textrm{P}_{c}^{+}$ may be
${\frac{3}{2}}^{-}$, ${\frac{5}{2}}^{-} $, ${\frac{3}{2}}^{+}$, ${\frac{5}{2}}^{+} $, independent of dynamical models.
\end{abstract}

\maketitle

\section{Introduction}

Searching for the multiquark states has been attracted great attentions for more than four decades~\cite{Jaffe:1977PRL,Jaffe:2005PR,Hicks:2015,Zhu:2016PR,Esposito:2017PR,Olsen:2018RMP,Guo:2018RMP}.
Analyzing the $J/\psi\,$p resonances in experiment at LHCb indicates that there exist pentaquark states including heavy quarks ($c$ and $\bar{c}$)~\cite{firstExpEvidence}.
One of the two resonances has a mass of $4380\pm8\pm29$ MeV and a width of $205\pm18\pm86$ MeV,
and another has a mass of $4449.8\pm1.7\pm2.5$ MeV and a width of $39\pm5\pm19$ MeV.
They are proposed to hold opposite parity, with one state having spin $3/2$ and the other $5/2$.
Even though quite a lot of theoretical investigations have been accomplished consequently
(see for example, Refs.~\cite{RChen:2015,HXChen:2015,Roca:2015,Mironov:2015,Guo:2015,Maiani:2015,He:2015,Zhao:2015a,Lebed:2015,Mikhasenko:2015,Meissner:2015,Anisovich:2015,Li:2015,Zhao:2015b,Ghosh:2015,Voloshin:2015,Riska:2015,Wang:2015a,Rosner:2015,Geng:2015,Wang:2015b}, for reviews, see Refs.~\cite{Hicks:2015,Zhu:2016PR} ),
the concrete spin and parity have not yet been well assigned to the two states.
Recently the pentaquark states are updated as $4312\,$MeV with a width about $9.8\,$MeV,
$4440\,$MeV with a width about $20.6\,$MeV and $4457\,$MeV with a width about $6.4\,$MeV~\cite{Aaij:2019PRL},  and the quantum numbers of the states have not yet been determined~\cite{Pilloni:2019PRL} either although many theoretical works have tried to shed light on that (for instance, by considering the states with molecular structure, the quantum numbers are proposed as $J^{P} = {\frac{1}{2}}^{-}, {\frac{1}{2}}^{-}, {\frac{3}{2}}^{-}$ (or ${\frac{5}{2}}^{-}$) (see, e.g., Refs.~\cite{Geng:2019PRL,Guo:2019PRD,Liu:2019PRD,Geng:2019PRD,Zhu:2019PRD,Zhu:2019PRD2,Zou:2019PRD,Sakai:2019PRD}),
or ${\frac{1}{2}}^{-}$(or ${\frac{3}{2}}^{-}$) and excited states with $J^{P}={\frac{1}{2}}^{-}$~\cite{Qiao:2019PLB}, which are not consistent with those proposed in Ref.~\cite{Winney:2019PRD}).
Furthermore, stringently speaking, pentaquark states refer to the color singlets composed of four quarks and one antiquark.
In such a compact pentaquark scenario, their quantum numbers are proposed as $J^{P} = {\frac{3}{2}}^{-}, {\frac{1}{2}}^{-}, {\frac{3}{2}}^{-}$ via considering the uncertainty and the rearrangement decay properties (see, e.g., Refs.~\cite{Cheng:2019PRD}).
Since fundamental calculation for the pentaquark states via QCD is not available at the present stage,
we investigate the quantum numbers by analyzing the symmetry, more explicitly the intrinsic nodal structure,
of the color singlets including four three flavor quarks and one antiquark in this paper.

The remainders of this paper are organized as follows.
In section \ref{se:SWF}, we give a classification of the wave functions of the pentaquark states according to the symmetries of the quarks.
Then in section \ref{se:nodalstructure} we assume that the four quarks form a tetrahedron or a square
%
%
and analyze the nodeless structure of the states with orbital angular momentum $L \leq 3$,
in turn, propose the quantum numbers of the pentaquark states.
Finally, we give our conclusion and some remarks in the last section.

\section{Structure of the Wave-functions }\label{se:SWF}

It has been known that the wave functions of few-body systems can usually be written as a coupling of the orbital part and an internal part, and the inherent nodal surface (INS) analysis approach is quite powerful to assign the quantum numbers to a system (see, e.g., Refs.~\cite{Bao:1997PRL,Bao:1999PRL,Liu:2002PLB,Liu:2003PRC}).
One can definitely take the INS analysis approach to investigate the pentaquark states ($q^{4}\bar{q}$ systems) at quark model level since they are typically few-body systems.
Since quark and antiquark can never identified with each other,
there does not exist antisymmetric restriction on the wave function when interchanging a quark with the antiquark.
However, if we consider only the four quarks, it should be antisymmetric, i.e., it has the symmetry $[1^{4}]$, if we ignore the current quark mass difference.
Considering the system that the four quarks involve three flavors (e.g., $uudc$, $uuds$, $uudb$, and so on), one knows well that the system possesses the internal symmetry
$SU_{CFS}(18) \supset SU_{C}(3) \otimes SU_{F}(3) \otimes SU_{S}(2)$.
Meanwhile the orbital part holds the symmetry of the permutation group $S_{4}$.
Let $[f]_{O}$, $[f]_{C}$ and $[f]_{FS}$ be the irreducible representation (IRREP) of the groups associated with the orbital, the color and the flavor-spin space, respectively, we should have

\begin{equation}
[1^4] \in [f]_{O} \otimes [f]_{C} \otimes [f]_{FS} \, .
\end{equation}
The lack of experimental observation of free ``color charge" indicates that
all the hadronic states should be SU(3) color-singlets.
For the $qqqq\bar{q}$ system, the IRREP of the SU(3) color space should be $[222]$.
Consequently, we have the IRREP of the color part of the four quark system as $[f]_{C} = [2 1 1]$.
Then the $[f]_O \otimes [f]_{FS}$ should be $[3\, 1]$, which is the conjugate of [211].
In turn, the configuration of the $[f]_{O}$ and $[f]_{FS}$ can be fixed with
the reduction principle of the direct product of the IRREPs of the unitary group.

The obtained possible $[f]_{FS}$ and the corresponding $[f]_O$ are listed in Table~\ref{tab:fo_ffs}.
Corresponding to each IRREP $[f]_{FS}$, the flavor and spin decomposition of the three flavors $q^4$ state and the $q^4\bar{q}$ state have been given in Ref.~\cite{Bijker:2004}.
We recall them here in Table~\ref{tab:fs_qqqq} and Table~\ref{tab:fs_qqqqqbar}.

It is obvious that such an orbital and flavor-spin configuration space is very large.
Therefore we should pick out the significant ones for the low-lying pentaquark states.
%

\begin{table}[htb]
\caption{\label{tab:fo_ffs} IRREPs of the flavor-spin symmetry corresponding to each possible orbital symmetry.}
\vspace*{2mm}
 \begin{tabular}{l|l}
 \hline
$[f]_O$ & $[f]_{FS}$\\
 \hline
 $[4]$  & $[31]$\\[-1.5mm]
 $[31]$ & $[4],\,[31],\,[22],\,[211]$\\[-1.5mm]
 $[22]$ & $[31],\,[211]$\\[-1.5mm]
 $[211]$ & $[31],\,[22],\,[211],\,[1111]$\\[-1.5mm]
 $[1111]$ & $[211]$\\
 \hline
\end{tabular}
\end{table}
\begin{table}[h]
\caption{\label{tab:fs_qqqq} Spin-flavor decomposition of the three flavor $q^4$ states (taken from Ref.~\cite{Bijker:2004}. The subscripts stand for the dimensions of the IRREP.).}
\vspace*{2mm}
\begin{tabular}{l|l|llll}
 \hline
$[f]_O$ & $SU_{FS}(6)$     & & $SU_{F}(3)$   & $\otimes$ & $SU_{S}(2)$\\
 \hline
$[4]$   & $[4]_{126}$   &  & $[4]_{15}$  & $\otimes$ & $[4]_5$ \\[-1.5mm]
	& 		&           & $[31]_{15}$ & $\otimes$ & $[31]_3$\\[-1.5mm]
	& 		&           & $[22]_{6}$  & $\otimes$ & $[22]_1$\\
 \hline
 $[31]$ & $[31]_{210}$  &           & $[4]_{15}$  & $\otimes$ & $[31]_3$\\[-1.5mm]
	& 		&           & $[31]_{15}$ & $\otimes$ & $[4]_5$ \\[-1.5mm]
	& 		&           & $[31]_{15}$ & $\otimes$ & $[31]_3$\\[-1.5mm]
	& 		&           & $[31]_{15}$ & $\otimes$ & $[22]_1$\\[-1.5mm]
	& 		&           & $[22]_{6}$  & $\otimes$ & $[31]_3$\\[-1.5mm]
	& 		&           & $[211]_{3}$ & $\otimes$ & $[22]_1$\\[-1.5mm]
	& 		&           & $[211]_{3}$ & $\otimes$ & $[31]_3$\\
 \hline
 $[22]$ & $[22]_{105}$  &           & $[4]_{15}$  & $\otimes$ & $[22]_1$\\[-1.5mm]
	& 		&           & $[31]_{15}$ & $\otimes$ & $[31]_3$\\[-1.5mm]
	& 		&           & $[22]_{6}$  & $\otimes$ & $[4]_5$ \\[-1.5mm]
	& 		&           & $[22]_{6}$  & $\otimes$ & $[22]_1$\\[-1.5mm]
	& 		&           & $[211]_{3}$ & $\otimes$ & $[31]_3$\\
 \hline
 $[211]$& $[211]_{105}$ &           & $[31]_{15}$ & $\otimes$ & $[31]_3$\\[-1.5mm]
	& 		&           & $[31]_{15}$ & $\otimes$ & $[22]_1$ \\[-1.5mm]
	& 		&           & $[22]_{6}$  & $\otimes$ & $[31]_3$\\[-1.5mm]
	& 		&           & $[211]_{3}$ & $\otimes$ & $[4]_5$\\[-1.5mm]
	& 		&           & $[211]_{3}$ & $\otimes$ & $[31]_3$\\[-1.5mm]
	& 		&           & $[211]_{3}$ & $\otimes$ & $[22]_1$\\
 \hline
$[1111]$& $[1111]_{15}$ &           & $[22]_{6}$  & $\otimes$ & $[22]_1$\\[-1.5mm]
	& 		&           & $[211]_{3}$ & $\otimes$ & $[31]_3$ \\
 \hline
\end{tabular}
\end{table}

\begin{table}[htb]
\caption{\label{tab:fs_qqqqqbar} Spin-flavor decomposition of the $q^4\bar{q}$ states (taken from Ref.~\cite{Bijker:2004}. The subscripts stand for the dimensions of the IRREP.).}
\vspace*{1mm}
\begin{tabular}{l|l|llll}
\hline
$[f]_O$ 	& $SU_{sf}(6)$   	&  & $SU_f(3)$   	& $\otimes$ & $SU_s(2)$	\\
%
%
%
%
\hline
$[4]$   	& $[51111]_{700}$   	&           & $[51]_{35}$  	& $\otimes$ & $[5]_6$ 	\\[-1mm]
		& 			&           & $[51]_{35}$  	& $\otimes$ & $[41]_4$	\\[-1mm]
		& 			&           & $[42]_{27}$  	& $\otimes$ & $[41]_4$	\\[-1mm]
		& 			&           & $[42]_{27}$  	& $\otimes$ & $[32]_2$	\\[-1mm]
		& 			&           & $[33]_{10}$  	& $\otimes$ & $[32]_2$	\\[-1mm]
		& 			&           & $[411]_{10}$ 	& $\otimes$ & $[5]_6$	\\[-1mm]
		& 			&           & $[411]_{10}$ 	& $\otimes$ & $[41]_4$	\\[-1mm]
		& 			&           & $[411]_{10}$ 	& $\otimes$ & $[32]_2$	\\[-1mm]
		& 			&           & $[321]_{8}$  	& $\otimes$ & $[41]_4$	\\[-1mm]
		& 			&           & $[321]_{8}$  	& $\otimes$ & $[32]_2$	\\
 \hline
$[4]+[31]$	& $[411111]_{56}$   	&           & $[411]_{10}$  	& $\otimes$ & $[41]_4$ \\[-1mm]
		& 			&           & $[321]_{8}$ 	& $\otimes$ & $[32]_2$ \\
 \hline
 $[31]$ 	& $[42111]_{1134}$  	&           & $[51]_{35}$  	& $\otimes$ & $[41]_4$\\[-1mm]
		& 		  	&           & $[51]_{35}$  	& $\otimes$ & $[32]_2$\\[-1mm]
 	 	& 		  	&           & $[42]_{27}$  	& $\otimes$ & $[5]_6$ \\[-1mm]
 	 	& 		  	&           & $2([42]_{27}$  	& $\otimes$ & $[41]_4)$\\[-1mm]
 	 	& 		  	&           & $2([42]_{27}$  	& $\otimes$ & $[32]_2)$\\[-1mm]
 	 	& 		  	&           & $[33]_{10}$  	& $\otimes$ & $[41]_4$\\[-1mm]
 	 	& 		  	&           & $[33]_{10}$  	& $\otimes$ & $[32]_2$\\[-1mm]
 	 	& 		  	&           & $[411]_{10}$  	& $\otimes$ & $[5]_6$\\[-1mm]
 		& 		  	&           & $2([411]_{10}$ 	& $\otimes$ & $[41]_4)$\\[-1mm]
 	 	& 		  	&           & $2([411]_{10}$ 	& $\otimes$ & $[32]_2)$\\[-1mm]
 	 	& 		  	&           & $[321]_{8}$  	& $\otimes$ & $[5]_6$\\[-1mm]
 	 	& 		  	&           & $2([321]_{8}$  	& $\otimes$ & $[41]_4)$\\[-1mm]
 	 	& 		  	&           & $2([321]_{8}$  	& $\otimes$ & $[32]_2)$\\[-1mm]
 	 	& 		  	&           & $[222]_{1}$  	& $\otimes$ & $[41]_4$\\[-1mm]
 	 	& 		  	&           & $[222]_{1}$  	& $\otimes$ & $[32]_2$\\
 \hline
\end{tabular}
\end{table}

\begin{center}
\begin{tabular}{l|l|llll}
\multicolumn{6}{r} (continuation of Table~\ref{tab:fs_qqqqqbar} ) \\
\hline
$[f]_O$ 	& $SU_{sf}(6)$   	&  & $SU_f(3)$   	& $\otimes$ & $SU_s(2)$	\\
%
\hline
%
$[31]+[22]+[211]$& $[321111]_{70}$   	&           & $[411]_{10}$  	& $\otimes$ & $[32]_2$ \\[-1mm]
		& 			&           & $[321]_{8}$ 	& $\otimes$ & $[41]_4$\\[-1mm]
		& 			&           & $[321]_{8}$  	& $\otimes$ & $[32]_2$\\[-1mm]
		& 			&           & $[222]_{1}$  	& $\otimes$ & $[32]_2$\\[-1mm]
 \hline
 $[22]$ 	& $[33111]_{560}$  	&           & $[51]_{35}$  	& $\otimes$ & $[32]_2$\\[-1mm]
 	 	& 		  	&           & $[42]_{27}$  	& $\otimes$ & $[41]_4$\\[-1mm]
 	 	& 		  	&           & $[42]_{27}$  	& $\otimes$ & $[32]_2$\\[-1mm]
 	 	& 		  	&           & $[33]_{10}$  	& $\otimes$ & $[5]_6$\\[-1mm]
 	 	& 		  	&           & $[33]_{10}$  	& $\otimes$ & $[41]_4$\\[-1mm]
 		& 		  	&           & $[33]_{10}$ 	& $\otimes$ & $[32]_2$\\[-1mm]
 	 	& 		  	&           & $[411]_{10}$ 	& $\otimes$ & $[41]_4$\\[-1mm]
 	 	& 		  	&           & $[411]_{10}$  	& $\otimes$ & $[32]_2$\\[-1mm]
 	 	& 		  	&           & $[321]_{8}$  	& $\otimes$ & $[5]_6$\\[-1mm]
 	 	& 		  	&           & $2([321]_{8}$  	& $\otimes$ & $[41]_4)$\\[-1mm]
 	 	& 		  	&           & $2([321]_{8}$  	& $\otimes$ & $[32]_2)$\\[-1mm]
 	 	& 		  	&           & $[222]_{1}$  	& $\otimes$ & $[41]_4$\\[-1mm]
 \hline
 $[211]$	& $[3211]_{540}$ 	&           & $[42]_{7}$ 	& $\otimes$ & $[41]_4$ \\[-1mm]
		& 			&           & $2([42]_{7}$ 	& $\otimes$ & $[32]_2$ \\[-1mm]
		& 			&           & $[33]_{10}$ 	& $\otimes$ & $[41]_4$ \\[-1mm]
		& 			&           & $[33]_{10}$ 	& $\otimes$ & $[32]_2$ \\[-1mm]
		& 			&           & $[411]_{10}$ 	& $\otimes$ & $[41]_4$ \\[-1mm]
		& 			&           & $[411]_{10}$ 	& $\otimes$ & $[32]_2$ \\[-1mm]
		& 			&           & $[321]_{8}$ 	& $\otimes$ & $[5]_6$ \\[-1mm]
		& 			&           & $2([321]_{8}$ 	& $\otimes$ & $[41]_4)$ \\[-1mm]
		& 			&           & $2([321]_{8}$ 	& $\otimes$ & $[32]_2)$ \\[-1mm]
		& 			&           & $[222]_{1}$ 	& $\otimes$ & $[5]_6$ \\[-1mm]
		& 			&           & $[222]_{1}$ 	& $\otimes$ & $[41]_4$ \\[-1mm]
		& 			&           & $[222]_{1}$ 	& $\otimes$ & $[32]_2$ \\[-1mm]
 \hline
$[211]+[1111]$  & $[222111]_{20}$   	&           & $[321]_{8}$  	& $\otimes$ & $[32]_2$ \\[-1mm]
		& 			&           & $[222]_{1}$ 	& $\otimes$ & $[41]_4$\\[-1mm]
 \hline
$[1111]$	& $[22221]_{70}$ 	&           & $[33]_{10}$  	& $\otimes$ & $[32]_2$\\[-1mm]
		& 			&           & $[321]_{8}$ 	& $\otimes$ & $[41]_4$\\[-1mm]
		& 			&           & $[321]_{8}$ 	& $\otimes$ & $[32]_2$\\[-1mm]
		& 			&           & $[222]_{1}$ 	& $\otimes$ & $[32]_2$\\
 \hline
\end{tabular}
\end{center}

\section{Inherent Nodal Structure Analysis}\label{se:nodalstructure}

From Tables~\ref{tab:fo_ffs}-\ref{tab:fs_qqqqqbar} we can notice that, given an IRREP of the orbital wave function, we can deduce the corresponding flavor and spin part of the wave function.
Generally, orbital wave function depends on the geometric configuration and the dynamics, or, in present stage, we should say models.
In this section, we analyze the structure of the wave functions of the pentaquark states independent of dynamical models.

The zero points for the wave-function are known as nodes.
The locus for the nodes is a surface ,i.e., the nodal surface, in configuration space.
For usual bound states of a quantum system,
the fewer nodes the configuration contains, the lower energy the state has.
The dynamical nodal surface depends on the dynamics of the system,
while the inherent nodal surface relies only on the inherent geometric configuration of the system.

Let $\mathcal{A}$ be a geometric configuration in the coordinate space of a quantum system, $\hat{O_i}$ be an element of the operation on the wave function $F(\mathcal{A})$,
we have
\begin{equation}
 \hat{O_i}F(\mathcal{A}) = F(\hat{O_i} \mathcal{A}).
\end{equation}
If the operation {$\hat{O_i}$} leaves the configuration $\mathcal{A}$ invariant, i.e., $\hat{O_i}\mathcal{A}=\mathcal{A}$, the above equation becomes
\begin{equation}\label{eq:nodalequation}
 \hat{O_i}F(\mathcal{A}) = F(\mathcal{A}).
\end{equation}
If there do not exist common non-zero solutions for the wave functions governed by Eq.~(\ref{eq:nodalequation}), all the $F(\mathcal{A})$ must be zero at the configuration $\mathcal{A}$, then a nodal surface appears.

The wave function of the four quarks with total angular momentum $J$ and parity $\pi$ can be written as
\begin{equation}
 \Psi = \sum_{L,S,\lambda} \Psi^S_{L\pi\lambda} \, ,
\end{equation}
with
\begin{equation}
 \Psi^S_{L\pi\lambda} = \sum_{i,M,M_{S}^{}}C^{J,M_{J}^{} }_{LM,SM_{S}^{} } F^{\pi\lambda_i}_{LM}\chi^{\tilde{\lambda_i}}_{S M_{S}^{}},
\end{equation}
where $F^{\pi\lambda_i}_{LM}$ is a function of the spatial coordinates, $\lambda$ denotes an  IRREP of the $S_{4}$ group, and $i$ index the basis vectors of that IRREP.
The $L$ and $S$ is the total orbital angular momentum and the total spin of the four quarks,  respectively. They couple to form $J$ via the reduction principle of $SU(2) \otimes SU(2) \supset SU(2)$, i.e., $C^{J,M_{J}^{} }_{LM,SM_{S}^{} }$ is the Clebsch-Gordan coefficient
where the $M$, $M_S$ and $M_J$ are the Z-components of $L$, $S$ and $J$, respectively.
The parity of the four-quark cluster is given by $\pi=(-1)^L$,
and the total parity of the pentaquark state $P=-\pi=(-1)^{L+1}$ since the antiquark holds an instinct negative parity.
$\chi^{\tilde{\lambda_i}}_{SM_S}$ is the wave function of the inner space, belonging to the $\tilde{\lambda_i}$-representation, the conjugate of $\lambda$.
$F^{\pi\lambda_i}_{LM}$ spans a representation of the direct product of the rotation group, the space inversion group and the permutation group $S_4$.
Thus the transformation property of the $F^{\pi\lambda_i}_{LM}$ with respect to the operations in terms of the above groups is prescribed, which will impose a very strong constraint on the $F^{\pi\lambda_i}_{LM}$.

\begin{figure}
  \includegraphics[width=0.45\textwidth]{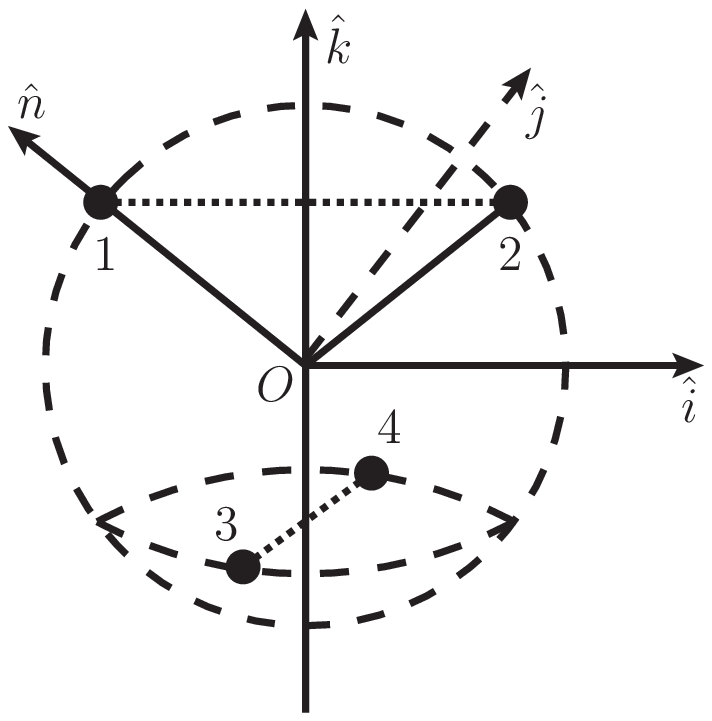}
  \includegraphics[width=0.45\textwidth]{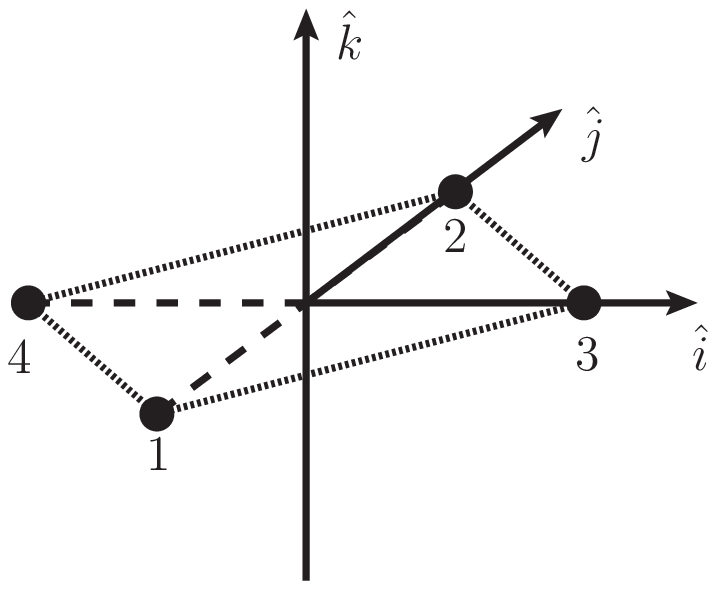}
 \caption{\label{fig:pentaquark} {\it Left panel:} body frame illustration of the equilateral tetradedron (ETH) configuration;  {\it Right panel:} body frame illustration of the square.}
\end{figure}

We assume that the four quarks form an equilateral tetrahedron(ETH) or a square, and the antiquark locates at the center of the four quark cluster, which is illustrated in Fig.\ref{fig:pentaquark}.
We determine all the accessible configuration with $L\leq 3$, extending the results of Ref.~\cite{Liu:2004}. For the case of equilateral tetrahedron, we denotes $O$ as the center of the four quarks, $\hat{i},\,\hat{j},\,\hat{k}$ form the frame, and $r_{12}\perp\hat{k}$, $r_{34}\perp\hat{k}$.
Referring to $R^{\vec{\nu}}_{\delta}$ as a rotation about the axis along the vector $\vec{\nu}$ by an angle $\delta$, $p_{ij}^{}$ as an interchange of the particles $i$ and $j$,
$p_{ijk}^{}$ and $p_{ijkl}^{}$ as permutation among the three and four particles, and $\hat{P}$ as a space inversion,  the ETH is evidently invariant to the operations:
\begin{eqnarray}\label{eq:nodaleq1}
 \hat{O}_{1} & = & p_{12}p_{34}R^{\hat{k}}_\pi,\\
 \hat{O}_{2} & = & p_{34}R^{\hat{j}}_\pi\hat{P},\\
 \hat{O}_{3} & = & p_{1423}R^{\hat{k}}_{\pi/2}\hat{P},\\\label{eq:nodaleq4}
 \hat{O}_{4} & = & p_{243}R^{\hat{n}}_{2\pi/3}.
 \end{eqnarray}
Inserting each of these operators to Eq.(\ref{eq:nodalequation}), we have a set of equations,
\begin{equation}\label{eq:nodal0eq}
 (\hat{O}_{i} - E) F^{\pi\lambda_j}_{LM}(\mathcal{A}) = 0 \, .
\end{equation}
$F^{\pi\lambda_j}_{LM}(\mathcal{A})$ is the orbital wave-function in the body frame, $E$ is the identity matrix, $j$ and $M$ denote the constituent of $\lambda$ and $L$.
Eqs.~(\ref{eq:nodaleq1})-(\ref{eq:nodaleq4}) are the equations that $F^{\pi\lambda_j}_{LM}(\mathcal{A})$ has to fulfill.
They are homogeneous linear algebraic equations. If they have common non-zero solutions, we have a nodeless structure.
The representations of the permutation group $S_{4}$ and the rotation group are represented in Table~\ref{tab:permutationelements} and Table~\ref{tab:rotationelements} where we just show those for $L=1$ as an example.
The equations can be solved easily.
The obtained results are showed in Table~\ref{tab:accessibility-OrbitalConfiguration}, where $A$ denotes the accessible configuration (holds nodeless structure) and ``$-$" stands for the inaccessible configuration.

\begin{table}
 \caption{\label{tab:permutationelements} IRREPs of the permutation group $S_4$ under the standard basis.}
  \begin{tabular}{|l|l|}
 \hline
 $[4]$	& $p_{12}=p_{23}=p_{34}=p_{1423}=p_{243}=p_{1324}=1$,\\
 \hline
$[31]$	& $p_{12}=\left(\begin{array}{ccc}-1 & 0 & 0\\-1 & 1 & 0\\-1 & 0 & 1\end{array}\right)$,
	  $p_{23}=\left(\begin{array}{ccc}0 & 1 & 0\\1 & 0 & 0\\0 & 0 & 1\end{array}\right)$,
	  $p_{34}=\left(\begin{array}{ccc}1 & 0 & 0\\0 & 0 & 1\\0 & 1 & 0\end{array}\right)$,\\
	& $p_{1423}=\left(\begin{array}{ccc}0 & 1 & -1\\0 & 0 & -1\\1 & 0 & -1\end{array}\right)$,
	  $p_{243}=\left(\begin{array}{ccc}0 & 0 & 1\\1 & 0 & 0\\0 & 1 & 0\end{array}\right)$,
	  $p_{1324}=\left(\begin{array}{ccc}0 & -1 & 1\\1 & -1 & 0\\0 & -1 & 0\end{array}\right)$,\\
 \hline
 $[22]$	& $p_{12}=\left(\begin{array}{cc}-1 & 0\\-1 & 1\end{array}\right)$,
	  $p_{23}=\left(\begin{array}{cc}0 & 1\\1 & 0\end{array}\right)$,
	  $p_{34}=\left(\begin{array}{cc}-1 & 0\\-1 & 1\end{array}\right)$,\\
	& $p_{1423}=\left(\begin{array}{cc}-1 & 0\\-1 & 1\end{array}\right)$,
	  $p_{243}=\left(\begin{array}{cc}-1 & 1\\-1 & 0\end{array}\right)$,
	  $p_{1324}=\left(\begin{array}{cc}-1 & 0\\-1 & 1\end{array}\right)$,\\
 \hline
$[211]$	& $p_{12}=\left(\begin{array}{ccc}1 & 0 & 0\\1 & -1 & 0\\1 & 0 & -1\end{array}\right)$,
	  $p_{23}=\left(\begin{array}{ccc}0 & -1 & 0\\-1 & 0 & 0\\0 & 0 & -1\end{array}\right)$,
	  $p_{34}=\left(\begin{array}{ccc}-1 & 0 & 0\\0 & 0 & -1\\0 & -1 & 0\end{array}\right)$,\\
	& $p_{1423}=\left(\begin{array}{ccc}0 & -1 & 1\\0 & 0 & 1\\-1 & 0 & 1\end{array}\right)$,
	  $p_{243}=\left(\begin{array}{ccc}0 & 0 & 1\\1 & 0 & 0\\0 & 1 & 0\end{array}\right)$,
	  $p_{1324}=\left(\begin{array}{ccc}0 & 1 & -1\\-1 & 1 & 0\\0 & -1 & 0\end{array}\right)$,\\
 \hline
 $[1111]$	& $p_{12}=p_{23}=p_{34}=p_{1423}=p_{1324}=-1$, $p_{243}=1$.\\\hline
\end{tabular}
\end{table}

\begin{table}
 \caption{\label{tab:rotationelements} IRREPs of the rotation group, $j$ is the angular moment, $m_1$ and $m_2$ denote the rows and columns of the matrix, $\alpha,\,\beta,\,\gamma$ is the Eulerian angular.}
\begin{tabular}{|l|l|}
\hline
       & $D^j_{m_1,m_2}(\alpha,\beta,\gamma)=\sum_k\frac{(-1)^k \sqrt{(j+m_1)! (j-m_1)! (j+m_2)! (j-m_2)!} \exp (-i \alpha  m_1-i \gamma  m_2) \sin ^{2 k-m_1+m_2}\left(\frac{\beta }{2}\right) \cos ^{2 j-2 k+m_1-m_2}\left(\frac{\beta }{2}\right)}{k! (j-k+m_1)! (j-k-m_2)! (k-m_1+m_2)!}$\\
\hline
 $L=1$ & $R^{\hat{k}}_\pi = D(0,0,\pi) = \left(\begin{array}{ccc}-1 & 0 & 0\\0 & 1 & 0\\0 & 0 & -1\end{array}\right)$,
	 $R^{\hat{j}}_\pi = D(0,\pi,0) = \left(\begin{array}{ccc}0 & 0 & 1\\0 & -1 & 0\\1 & 0 & 0\end{array}\right)$,
         $R^{\hat{k}}_{\pi/2} = D(0,0,\pi/2) = \left(\begin{array}{ccc}-i & 0 & 0\\0 & 1 & 0\\0 & 0 & i\end{array}\right)$,\\
       & $R^{\hat{n}}_{2\pi/3} = D(0,-\theta,0)\cdot D(2\pi/3,\theta,0) = \left(\begin{array}{ccc}-i/2 & 1/2+i/2 & -1/2\\1/2+i/2 & 0 & -1/2+i/2\\-1/2 & -1/2+i/2 & i/2\end{array}\right)$, $\theta=\text{Arccos}(\frac{1}{\sqrt{3}})$.\\\hline
\end{tabular}
\end{table}

\begin{table}[!ht]
\caption{\label{tab:accessibility-OrbitalConfiguration} Obtained accessibility of the ETH configuration and the square configuration to the $(L\pi\lambda)$ and related configurations of the wave-functions.}
\vspace*{2mm}
\begin{tabular}{c|c|ccccc}
 \hline
	& ~$L^{\pi}$~ & $\;[4]\;$ & $\;[31]\;$ & $\;[22]\;$& $\;[211]\;$ & $\;[1111]\;$\\\hline
ETH	& $0^{+}$ & A	  & --	  & --	  & A	   & --	     \\
	& $1^{-}$ & --	  & A	  & A	  & A	   & A	     \\
	& $2^{+}$ & A	  & A	  & A	  & A    & A	     \\
	& $3^{-}$ & A	  & A	  & A   & A	   & A	     \\\hline
ETH$_3$	& $0^{+}$ & A	  & --	  & A	  & A	   & --	     \\
	& $1^{-}$ & --	  & A	  & A	  & A	   & A	     \\
	& $2^{+}$ & A	  & A	  & A	  & A	   & A	     \\
	& $3^{-}$ & A	  & A	  & A	  & A	   & A	     \\\hline
Square	& $0^{+}$ & A	  & --	  & A	  & --	   & --	     \\
	& $1^{-}$ & --	  & A	  & --	  & A   & --	     \\
	& $2^{+}$ & A	  & A	  & A	  & A	   & A	     \\
	& $3^{-}$ & --	  & A	  & --	  & A	   & --	     \\\hline
Square$_3$& $0^{+}$ & A	  & A	  & A	  & A	   & --	     \\
	& $1^{-}$ & --	  & A	  & --	  & A	   & --	     \\
	& $2^{+}$ & A	  & A   & A	  & A	   & A	     \\
	& $3^{-}$ & --	  & A	  & --	  & A	   & --	     \\\hline
\end{tabular}
\end{table}

It is believed that the wave functions are strongly constrained at the ETH configuration, however they are less constrained in the neighborhood of the ETH, since the ETH holds the highest symmetry.
To check such an idea, we have performed calculations in the case that the shape in the left panel of Fig.~\ref{fig:pentaquark} is invariant under the operations $\hat{O}_{1}$, $\hat{O}_{2}$ and $\hat{O}_{3}$, but not $\hat{O}_{4}$.
The obtained results are listed as those marked with ETH$_3$ in Table~\ref{tab:accessibility-OrbitalConfiguration}.
It shows apparently that all the configurations accessible to the ETH are definitely accessible to the ETH$_{3}$, and more nodeless configurations are allowed for the ETH$_{3}$.  .

For the case of the four quarks forming a square, the operations for the configuration to be invariant are
\begin{eqnarray}\label{eq:nodaleqsq1}
 \hat{O}_{1} & = & p_{12}p_{34}\hat{P},\\
 \hat{O}_{2} & = & R^{\hat{k}}_\pi\hat{P},\\
 \hat{O}_{3} & = & p_{34}R^{\hat{j}}_{\pi},\\\label{eq:nodaleqsq4}
 \hat{O}_{4} & = & p_{1324}R^{\hat{k}}_{\pi/2}.
 \end{eqnarray}
Inserting them into Eq.~(\ref{eq:nodalequation}), we get another set of equations as Eq.~(\ref{eq:nodal0eq}). The obtained accessible configuration of the $(L\pi\lambda)$ and the IRREPs of the $S_{4}$ group of the wave functions are listed as those marked as Square in Table~\ref{tab:accessibility-OrbitalConfiguration}.
In the neighborhood of the square, for example, prolonged along the $\hat{j}$ direction, it is still invariant under $\hat{O}_{1}$, $\hat{O}_{2}$ and $\hat{O}_{3}$, but not $\hat{O}_{4}$.
The accessible configurations are also given in Table~\ref{tab:accessibility-OrbitalConfiguration} with a mark Square$_3$.
It is evident that the square-accessible component is inherently nodeless in the neighborhood of the square.

Combining Table~\ref{tab:fs_qqqq} and Table~\ref{tab:accessibility-OrbitalConfiguration} one can get the number of the accessible (nodeless) states $J_{4}^{\pi} = L^{\pi} \oplus S$ with orbital angular momentum $L \leq 3$. The obtained result is listed in Table~\ref{tab:accessibility-AngularMomenta}.

\begin{table}[!ht]
\caption{\label{tab:accessibility-AngularMomenta} Obtained number of the accessible (nodeless) states $J^{P}$ with orbital angular momentum $L \leq 3$ of the ETH configuration and square configuration of the four quark system.}
\vspace*{2mm}
\begin{tabular}{c|cccccccccccc}
 \hline
Configuration & ~~$ 0^{+}$~ & ~~$0^{-}$~ & ~~$1^{+}$~ & ~~$1^{-}$~ & ~~$2^{+}$~ & ~~$2^{-}$~ & ~~$3^{+}$~ & ~~$3^{-}$~ & ~~$4^{+}$~ & ~~$4^{-}$~ & ~~$5^{+}$~ & ~~$5^{-}$~ \\\hline
ETH	& 7 & 10 & 19 & 22 & 25 & 28 & 15 & 26 & 4 & 15 & 0 & 4   \\
square & 7 & 7 & 18 & 15 & 25 & 19 & 15 & 17 & 4 & 10 & 0 & 2   \\
ETH$_{3}$ & 9 & 10 & 21 & 22 & 26 & 28 & 15 & 26 & 4 & 15 & 0 & 4   \\
square$_{3}$ & 11 & 7 & 25 & 15 & 27 & 19 & 15 & 17 & 4 & 10 & 0 & 2   \\
\hline
\end{tabular}
\end{table}

Considering the fact that, for usual bound states of quantum system,
the fewer nodes the configuration contains, the lower energy the state has.
One can recognize that the $J^{\pi}$ of the lowest energy state involving four quarks
with three flavors is $J_{4}^{\pi}=2^{+}$ (the accessible number is 50).
While that of the second, the third lowest state is $2^{-}$, $3^{-}$ (the accessible number is 47, 43, respectively).
If the symmetry of the configurations are less constrained, such as those for $\textrm{ETH}_{3}$ and $\textrm{square}_{3}$,
the result almost maintains the same (the accessible number becomes 53, 47, 46, respectively,
but the $J_{4}^{\pi}$ of the third lowest changes to $1^{+}$).

In our scheme regarding pentaquark states, we notice that the P$_{c}^{+}$ states are composed of a four quark cluster involving a $c$-quark ($uudc$) and an antiquark $\bar{c}$ at the ground state.
The total spin and the parity of the P$_{c}^{+}$ states should be $J^{P}(\textrm{P}_{c}^{+} ) = J_{4}^{\pi}(uudc) \oplus {\frac{1}{2}}^{-}(\bar{c}) \in \{ {\frac{3}{2}}^{-}, {\frac{5}{2}}^{-}, {\frac{3}{2}}^{+}, {\frac{5}{2}}^{+} \}$.
Since the $c$-quark is much heavier than the $u$- and $d$-quarks,
we can assume that the interaction between the cluster and the $\bar{c}$-quark is similar to that between
the $c$- and $\bar{c}$-quarks.
In view of heavy quark interaction, the energy sequence (from lower to higher) of the doublet with mainly $J_{4}^{\pi}=2^{\pm}$, is
$\{ {\frac{3}{2}}^{-}, {\frac{5}{2}}^{-} \}$, $\{ {\frac{3}{2}}^{+}, {\frac{5}{2}}^{+} \}$, respectively.
Taking the fact that the spin--orbital splitting is proportional to the orbital angular momentum into consideration,
one can assign the energy sequence of the $J^{P}$ states to ${\frac{3}{2}}^{-}$, ${\frac{5}{2}}^{-} $, ${\frac{3}{2}}^{+}$, ${\frac{5}{2}}^{+} $.

\section{Summary and Remarks}

In summary, by analyzing the symmetry, more concretely the inherent nodal structure, of
the system including four quarks involing three flavors,
we propose dynamical model independently that the quantum numbers of
the lowest-lying pentaquark states $\textrm{P}_{c}^{+}$ could be
$J^{P}=\frac{3}{2}^{-}$ with orbital angular momentum $L=2$ mainly.
The quantum numbers of the other two low-lying pentaquark states $\textrm{P}_{c}$ could be
$J^{P}= {\frac{3}{2}}^{+}$, ${\frac{5}{2}}^{+} $.
And there may exist another $\textrm{P}_{c}$ pentaquark state with $J^{P} = {\frac{5}{2}}^{-}$,
which may be observed by analyzing the structure of the peak around $4312\,$MeV.
More experiments or analyzing the existing experiments with more broad viewpoints
(see, e.g., Refs.~\cite{Voloshin:2019PRD,Chen:2019PLB}) are then expected.

It is finally remarkable that our present result is only valid for the pentaquark states without molecular structure,
namely no color singlet three-quark states and meson-like states.
Of course, such a result is only a qualitative result.
However, taking the presently assigned configuration into dynamical model
calculations can apparently help to reduce the load of numerical calculation tasks.
In addition, the work analyzing the inherent nodal structure of the pentaquark states with molecular configurations will be given in a sequel.

\medskip

\noindent
This work was supported by the National Natural Science Foundation of China under Contracts No.\ 11435001,
No. 11775041 , and the National Key Basic Research Program of China under Contract No.\ 2015CB856900.


\begin{thebibliography}{99}
\bibitem{Jaffe:1977PRL}
     R. L. Jaffe,
      {\it Perhaps a stable dihyperon},
        Phys. Rev. Lett. 38, 195 (1977).


\bibitem{Jaffe:2005PR}
      R. L. Jaffe,
      {\it Exotica},
        Phys. Rept. {\bf 409}, 1 (2005).

\bibitem{Hicks:2015}
      K. Hicks,   {\it  Elusive Pentaquark Comes into View },
        Phys. {\bf 8}, 77 (2015).

\bibitem{Zhu:2016PR}
      H.X. Chen, W. Chen, X. Liu, and S. L. Zhu,
      {\it The hidden-charm pentaquark and tetraquark states},
        Phys. Rept. {\bf 639}, 1 (2016);
      Y.R. Liu, H.X. Chen, W. Chen, X. Liu, and S. L. Zhu,
      {\it Pentaquark and tetraquark states},
        Prog. Part. Nucl. Phys. {\bf 107}, 237 (2019).

\bibitem{Esposito:2017PR}
      A. Esposito, A. Pilloni, and A. D. Plosa,
      {\it  Multiquark resonances},
        Phys. Rept. {\bf 668}, 1 (2017).

\bibitem{Olsen:2018RMP}
      S. L. Olsen, T. Skwarnicki, and D. Zieminska,
      {\it Nonstandard heavy mesons and baryons: Experimental evidence},
        Rev. Mod. Phys. {\bf 90}, 015003 (2018).

\bibitem{Guo:2018RMP}
      F.K. Guo, C. Hanhart, U.-G. Meissner, Q. Wang, Q. Zhao, and B. S. Zou,
      {\it Hadronic molecules},
        Rev. Mod. Phys. {\bf 90}, 015004 (2018).

\bibitem{firstExpEvidence}
      R. Aaij, {\it et al.} (LHCb collaboration),
      {\it Observation of $J/\psi\,$p resonances consistent with pentaquark states in $\Lambda^{0}_{b} \rightarrow J/\psi K^{-} p$ decays},
         Phys. Rev. Lett. {\bf 115}, 072001 (2015).

\bibitem{RChen:2015}
     R. Chen, X. Liu, X. Q. Li,  and S. L. Zhu,
      {\it Identifying exotic hidden-charm pentaquarks },
       Phys. Rev. Lett. {\bf 115}, 132002 (2015).

\bibitem{HXChen:2015}
     H. X. Chen, W. Chen, X. Liu, T. G. Steele, and S. L. Zhu,
      {\it Towards exotic hidden-charm pentaquarks in QCD},
       Phys. Rev. Lett. {\bf 115}, 172001 (2015).

\bibitem{Voloshin:2015}
     V. Kubarovsky, and M. B. Voloshin,
      {\it Formation of hidden-charm pentaquarks in photon-nucleon collisions},
        Phys. Rev. D {\bf 92}, 031502(R) (2015).

\bibitem{Zhao:2015b}
     Q. Wang, X. H. Liu, and Q. Zhao,
      {\it Photoproduction of hidden charm pentaquark states $P^{+}_{c}$(4380) and $P^{+}_{c}(4450)$},
       Phys. Rev. D {\bf 92}. 034022 (2015).

\bibitem{Riska:2015}
     N. N. Scoccola, D. O. Riska, and M. Rho,
      {\it On the pentaquark candidates $P_{c}^{+}$(4380) and $P_{c}^{+}$(4450) within the soliton picture of baryons},
       Phys. Rev. D {\bf 92}, 051501(R) (2015).

\bibitem{Guo:2015}
      F. K. Guo, U. -G. Meissner, W. Wang, and Z. Yang,
       {\it  How to reveal the exotic nature of the $P_{c}$(4450) },
        Phys. Rev. D {\bf 92}, 071502(R) (2015).

\bibitem{Roca:2015}
     L. Roca, J. Nieves, and E. Oset,
      {\it The LHCb pentaquark as a $D^{-*}\Sigma c\, - \, D^{-*} \Sigma^{*}c$ molecular state},
       Phys. Rev. D {\bf 92}, 094003 (2015).

\bibitem{Mironov:2015}
     A. Mironov, and A. Morozov,
      {\it Is pentaquark doublet a hadronic molecule? }
       JETP Lett. {\bf 102}, 271 (2015).

\bibitem{Maiani:2015}
     L. Maiani, A. D. Polosa, and V. Riquer,
       {\it The New Pentaquarks in the Diquark Model },
        Phys. Lett. {\bf B 749}, 289 (2015).

\bibitem{Lebed:2015}
     R. F. Lebed,
      {\it  The Pentaquark Candidates in the Dynamical Diquark Picture},
       Phys. Lett. B {\bf 749}, 454 (2015).

\bibitem{Meissner:2015}
     U. -G. Meissner, and J. A. Oller,
      {\it Testing the $\chi_{c1}$ p composite nature of the $P_{c}$(4450) },
        Phys. Lett. B {\bf 751}, 59 (2015).

\bibitem{Geng:2015}
     Y. K. Hsiao, and C. Q. Geng,
      {\it Pentaquarks from intrinsic charms in $\Lambda_{b}$ decays},
       Phys. Lett. B {\bf 751}, 572 (2015).

\bibitem{Li:2015}
     G. N. Li, X. G. He, and M. He,
      {\it Some Predictions of Diquark Model for Hidden Charm Pentaquark Discovered at the LHCb},
        JHEP {\bf 1512}, 128 (2015).

\bibitem{Rosner:2015}
     M. Karliner, and J. L. Rosner,
      {\it Photoproduction of Exotic Baryon Resonances},
         Phys. Lett. B {\bf 752}, 329 (2016).

\bibitem{He:2015}
     J. He,
      {\it The $D^{-}\Sigma^{*}c$ and $D^{-*}\Sigma c$ interactions and the LHCb hidden-charmed pentaquarks},
       Phys. Lett B {\bf 753}, 547 (2016)

\bibitem{Zhao:2015a}
     X. H. Liu, Q. Wang, and Q. Zhao,
      {\it  Understanding the newly observed heavy pentaquark candidates},
       Phys. Lett. B {\bf 757}, 231 (2016).

\bibitem{Mikhasenko:2015}
     M. Mikhasenko,
      {\it  A triangle singularity and the LHCb pentaquarks} ,
       arXiv:1507.06552.

\bibitem{Anisovich:2015}
     V. V. Anisovich, M. A. Matveev, J. Nyiri, A. V. Sarantsev, and A. N. Semenova,
       {\it Pentaquarks and resonances in the pJ/$\psi$ spectrum},
        arXiv:1507.07652;

\bibitem{Wang:2015a}
     Z. G. Wang,
      {\it Analysis of the $P_{c}$(4380) and $P_{c}$(4450) as pentaquark states in the diquark model with QCD sum rules},
       Eur. Phys. J. C {\bf 76}, 70 (2016).

\bibitem{Wang:2015b}
     Z. G. Wang, and T. Huang,
      {\it Analysis of the pentaquark states in the diquark model with QCD sum rules},
       Eur. Phys. J. C {\bf 76}, 43 (2016).

\bibitem{Ghosh:2015}
     R. Ghosh, A. Bhattacharya, and B. Chakrabarti,
      {\it The masses of $P_{c}^{+}$(4380) and $P_{c}^{+}$(4450) in the quasi particle diquark model},
        Phys. Part. Nucl. Lett. {\bf 14}, 550 (2017).

\bibitem{Aaij:2019PRL}
      R. Aaij, {\it et al.} (LHCb collaboration),
        {\it Observation of a Narrow Pentaquark State, $P_{c}^{+}(4312)$, and of the Two-peak Structure of the $P_{c}(4450)^{+}$},
         Phys. Rev. Lett. {\bf 122}, 222001 (2019).

\bibitem{Pilloni:2019PRL}
     C. Fern\'{a}ndez-Ram\'{i}rez, A. Pilloni, M. Albaladejo, A. Jackura, V. Mathieu, M. Mikhasenko, J. A. Silva-Castro, and A. P. Szczepaniak,
      {\it Interpretation of the LHCb $P_{c}^{+}(4312)$ Signal},
        Phys. Rev. Lett. {\bf 123}, 092001 (2019).

\bibitem{Geng:2019PRL}
      M. Z. Liu, Y. W. Pan, F. Z. Peng, M. S. S\'{a}nchez, and L. S. Geng,
       {\it Emergence of a Complete Heavy-quark SpinSymmetry Multiplet: Seven Molecular Pentaquarks in Light of the Latest LHCb Analysis},
         Phys. Rev. Lett. {\bf 122}, 242001 (2019).

\bibitem{Guo:2019PRD}
      F. K. Guo, H. J. Jing, U.-G. Meissner, and S. Sakai,
       {\it Isospin breaking decays as a diagnosis of the hadronic molecular structure of the $P_{c}(4457)$},
         Phys. Rev. D {\bf 99}, 091501(R) (2019).

\bibitem{Liu:2019PRD}
      R. Chen, Z. F. Sun, X. Liu, and S. L. Zhu,
       {\it Strong LHCb evidence supporting the existence of the hidden-charm molecular pentaquarks},
         Phys. Rev. D {\bf 100}, 011502(R) (2019).

\bibitem{Geng:2019PRD}
      C. J. Xiao, Y. Huang, Y. B. Dong, L. S. Geng, and D. Y. Chen,
       {\it Exploring the molecular scenario $P_{c}(4312)$, $P_{c}(4440)$, and $P_{c}(4457)$},
         Phys. Rev. D {\bf 100}, 014022 (2019).

\bibitem{Zhu:2019PRD}
     L. Meng, B. Wang, G. J. Wang, and S. L. Zhu,
      {\it Hidedn charm pentaquark states and $\Sigma_{c}\bar{D}^{(*)}$ interaction in chiral perturbation theory},
        Phys. Rev. D {\bf 100}, 014031 (2019).

\bibitem{Zhu:2019PRD2}
     H. X. Zhu, W. Chen, and S. L. Zhu,
      {\it Possible interpretation of the $P_{c}(4312)$, $P_{c}(4440)$, and $P_{c}(4457)$},
        Phys. Rev. D {\bf 100}, 051501(R) (2019).

\bibitem{Zou:2019PRD}
     Y. H. Liu, and B. S. Zou,
      {\it Strong decays of the latest LHCb pentaquark candidates in hadronic molecule picture},
        Phys. Rev. D {\bf 100}, 056005 (2019).

\bibitem{Sakai:2019PRD}
     S. Sakai, H. J. Jing, and F. K. Guo,
      {\it Decays of $P_{c}$ into $J/\psi N$ and $\eta_{c}N$ with heavy quark spin symmetry},
        Phys. Rev. D {\bf 100}, 074007 (2019).

\bibitem{Qiao:2019PLB}
     R.L. Zhu, X.J. Liu, H.X. Huang, and C. F. Qiao,
      {\it Analyzing doubly heavy tetra- and penta-quark states by vibrational method},
        Phys. Lett. B {\bf 797}, 134869 (2019).

\bibitem{Winney:2019PRD}
     D. Winney, C. Fanelli, A. Pilloni, A. N. Hiller Blin, C. Fern\'{a}ndez-Ram\'{i}rez, M. Albaladejo,
       V. Mathieu, V. I. Mokeev, and A. P. Szczepaniak,
      {\it Double polarization observables in pentaquark photoproduction},
        Phys. Rev. D {\bf 100}, 034019 (2019).

\bibitem{Cheng:2019PRD}
     J. B. Cheng, and Y. R. Liu,
      {\it $P_{c}(4457)^{+}$, $P_{c}(4440)^{+}$ and $P_{c}(4312)^{+}$: Molecules or compact pentaquarks},
        Phys. Rev. D {\bf 100}, 054002 (2019).

\bibitem{Bao:1997PRL}
     C. G. Bao, and C. D. Lin,
      {\it Effect of quantum-mechnical symmetry on the geometric structures of intershell states of three-valence-electron atoms},
        Phys. Rev. A {\bf 52}, 3586 (1995).
     C. G. Bao,
      {\it Large regions of stability in the phase diagrams of quantum dots and the associated filling factors},
        Phys. Rev. Lett. {\bf 79}, 3475 (1997).

\bibitem{Bao:1999PRL}
     C. G. Bao, and Y. X. Liu,
      {\it Deduction of the quantum numbers of low-lying states of 6-nucleon systems based on symmetry},
        Phys. Rev. Lett. {\bf 82}, 61 (1999).

\bibitem{Liu:2002PLB}
     Y. X. Liu, J. S. Li, and C. G. Bao,
      {\it Candidates of low-lying dibaryons with strangeness $S\leq 5$ in a symmetry analysis},
        Phys. Lett. B {\bf 544}, 280 (2002).

\bibitem{Liu:2003PRC}
     Y. X. Liu, J. S. Li, and C. G. Bao,
      {\it Low-lying S-wave and P-wave dibaryons in a nodal structure analysis},
        Phys. Rev. C {\bf 67}, 055207 (2003).


\bibitem{Bijker:2004}
     R. Bijker, M. M. Giannini, and E. Santopinto,
      {\it Spectroscopy of pentaquark states},
        Eur. Phys. J. {\bf A 22}, 319 (2004).

\bibitem{Liu:2004}
     Y. X, Liu, J. S. Li, and C. G. Bao,
      {\it Low-lying $s=+1$ Pentaquark States in the Inherent Nodal Structure Analysis},
         arXiv:hep-ph/0401197

\bibitem{Voloshin:2019PRD}
     M. B. Voloshin,
      {\it Some decay properties of hidden-charm pentaquarks as baryon-meson mocecules},
        Phys. Rev. D {\bf 100}, 034020 (2019).

\bibitem{Chen:2019PLB}
     X. Y. Wang, J. He, X. R. Chen, Q.J. Wang, and X.M. Zhu,
      {\it Pion-induced production of hidden-charm pentaquarks $P_{c}(4312)$, $P_{c}(4440)$, and $P_{c}(4457)$},
        Phys. Lett. B {\bf 797}, 134869 (2019).
\end{thebibliography}
\end{document}